\documentclass[12pt]{article}
\usepackage{a4wide}
\usepackage{latexsym}
\usepackage{axodraw}
\usepackage{amsmath}
\usepackage{amsfonts}
\usepackage{amscd}
\usepackage{cite}

\usepackage{pslatex}
\usepackage{graphicx}
\usepackage[latin1]{inputenc}
\usepackage[T1]{fontenc}

\newcommand{\bq}{\begin{eqnarray}}
\newcommand{\eq}{\end{eqnarray}}
\newcommand{\eps}{\varepsilon}

\newcommand{\R}{\mathbb{R}}
\newcommand{\C}{\mathbb{C}}
\newcommand{\Z}{\mathbb{Z}}

\newtheorem{satz}{}

\newtheorem{Prop}[satz]{Proposition}
\newtheorem{Lemma}[satz]{Lemma}

\begin{document}

\thispagestyle{empty}

\begin{flushright}
  MZ-TH/11-48 
\end{flushright}

\vspace{1.5cm}

\begin{center}
  {\Large\bf A second-order differential equation for the two-loop sunrise graph with arbitrary masses\\
  }
  \vspace{1cm}
  {\large Stefan M\"uller-Stach ${}^{a}$, Stefan Weinzierl ${}^{b}$ and Raphael Zayadeh ${}^{a}$\\
  \vspace{1cm}
      {\small ${}^{a}$ \em Institut f{\"u}r Mathematik, Universit{\"a}t Mainz,}\\
      {\small \em D - 55099 Mainz, Germany}\\
  \vspace{2mm}
      {\small ${}^{b}$ \em Institut f{\"u}r Physik, Universit{\"a}t Mainz,}\\
      {\small \em D - 55099 Mainz, Germany}\\
  } 
\end{center}

\vspace{2cm}

\begin{abstract}\noindent
  {
We derive a second-order differential equation for the two-loop sunrise graph in two dimensions with arbitrary masses.
The differential equation is obtained by viewing the Feynman integral as a period of a variation
of a mixed Hodge structure, where the variation is with respect to the external momentum squared.
The fibre is the complement of an elliptic curve.
From the fact that the first cohomology group of this elliptic curve is two-dimensional we obtain
a second-order differential equation.
This is an improvement compared to the usual way of deriving differential equations:
Integration-by-parts identities lead only to a coupled system of four first-order differential equations.
   }
\end{abstract}

\vspace*{\fill}

\newpage

\section{Introduction}
\label{sec:intro}

Precision calculations in high energy particle physics require the computation of quantum corrections.
These can be visualised by Feynman loop diagrams. In this paper we consider a particular loop diagram,
the two-loop sunrise graph, shown in fig.~(\ref{fig_sunrise_graph}).
This graph has received in the past significant attention 
in the literature \cite{Caffo:1998du,Davydychev:1999ic,Caffo:1999nk,Caffo:2001de,Onishchenko:2002ri,Argeri:2002wz,Caffo:2002ch,Czyz:2002re,Laporta:2004rb,Pozzorini:2005ff,Caffo:2008aw,Groote:2005ay}.
Despite this effort, an analytical answer in the general case of unequal masses is not yet known.
The state-of-the-art for the two-loop sunrise graph can be summarised as follows:
In the special case where all three internal masses are equal, 
a second-order differential equation in the external momentum squared and its analytical solution are known \cite{Laporta:2004rb}.
In the general case of unequal masses integration-by-parts identities \cite{Tkachov:1981wb,Chetyrkin:1981qh}
can be used to relate integrals with different powers of the propagators.
In the case of the sunrise topology with unequal masses all integrals can be expressed in terms of four master integrals plus simpler integrals.
This results in a coupled system of four first-order differential equations for the four master integrals \cite{Caffo:1998du}.
For practical applications this system can be solved numerically \cite{Caffo:2002ch,Caffo:2008aw}.

In this paper we reconsider the two-loop sunrise graph in two dimensions with unequal masses. 
We will show that also in the unequal mass case there is a second-order differential equation for a single master integral.
(The other three master integrals are then given as the derivatives with respect to the three internal masses.)
This second-order differential equation is not obtained from integration-by-parts identities.
Setting all masses equal, the second-order differential equation agrees with the well-known second-order differential equation for the equal
mass case.

How is this second-order differential equation obtained? 
Our method is interesting in its own right and not limited to the special case of the two-loop sunrise diagram.
We expect the method to be applicable to other loop integrals as well.
The starting point is the relation between Feynman integrals 
and periods of motives \cite{Bloch:2006,Bloch:2008jk,Aluffi:2008sy,Aluffi:2008rw,Aluffi:2009b,Marcolli:2009a,Aluffi:2009a,Rej:2009ik,Ceyhan:2010a,Belkale:2003,Bogner:2007mn}.
We view the two-loop sunrise integral as a period of a variation of a mixed Hodge structure.
The variation is with respect to the external momentum squared.
In the case of the two-loop sunrise integral we find that the fibre is the complement of an elliptic curve.
The theory of elliptic curves is well studied.
In particular, there is a second-order differential equation -- the Picard--Fuchs equation -- related to a family of elliptic curves.
The Picard--Fuchs equation expresses the fact that the first cohomology groups of the elliptic curves are two-dimensional.
The groups are generated by the holomorphic one-form and its first derivative with respect to the variation parameter.
It follows that the second derivative must be a linear combination of these two generators, which leads to a
second-order differential equation.
This is the the sought-after differential equation for the two-loop sunrise graph.

In this paper we give a detailed account how the second-order differential equation is obtained.
The analytic solution of this equation is beyond the scope of the present paper and will be dealt with in a future publication.
We expect that a solution can be obtained along the lines of refs.~\cite{Laporta:2004rb,Argeri:2007up}.

This paper is organised as follows:
In the next section we define the two-loop sunrise integral and recall a few basic facts.
Section~\ref{sec:derivation} is devoted to the derivation of the second-order differential equation.
This section is divided into several subsections. 
We start with presenting the formalism in subsection~\ref{subsec:formalism}.
The concrete calculations are split into two parts.
In the first step we derive the coefficients of the homogeneous part of the equation in subsection~\ref{subsec:homogeneous}.
In a subsequent step we obtain the inhomogeneous terms. This is done in subsection~\ref{subsec:inhomogeneous}.
For the convenience of the reader the complete result is summarised in subsection~\ref{subsec:result}.
Finally, our conclusions are contained in section~\ref{sec:conclusions}.
In an appendix we discuss the relations between the two-loop sunrise integrals in $D=2$ and $D=4-2\eps$ dimensions.

\section{Definition of the two-loop sunrise integral}
\label{sec:definition}

\begin{figure}
\begin{center}
\begin{picture}(100,100)(0,0)
\Vertex(20,50){2}
\Vertex(80,50){2}
\Line(20,50)(80,50)
\Line(80,50)(100,50)
\Line(0,50)(20,50)
\CArc(50,50)(30,0,180)
\CArc(50,50)(30,180,360)
\Text(105,50)[l]{$p$}
\Text(50,85)[b]{$m_1$}
\Text(50,55)[b]{$m_2$}
\Text(50,25)[b]{$m_3$}
\end{picture}
\end{center}
\caption{
The two-loop sunrise graph.
}
\label{fig_sunrise_graph}
\end{figure}
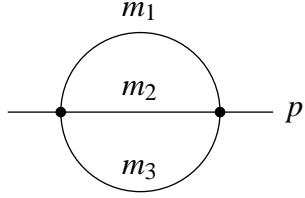

The two-loop integral corresponding to the sunrise graph with arbitrary masses is given 
in $D$-dimensional Minkowski space by
\bq
\label{def_sunrise}
\lefteqn{
 S\left( D, p^2, m_1^2, m_2^2, m_3^2, \mu^2 \right)
 = } & &
 \nonumber \\
 & &
 \left(\mu^2\right)^{3-D}
 \int \frac{d^Dk_1}{i \pi^{\frac{D}{2}}} \frac{d^Dk_2}{i \pi^{\frac{D}{2}}}
 \frac{1}{\left(-k_1^2+m_1^2\right)\left(-k_2^2+m_2^2\right)\left(-\left(p-k_1-k_2\right)^2+m_3^2\right)}.
\eq
The corresponding sunrise graph is shown in fig.~(\ref{fig_sunrise_graph}).
In eq.~(\ref{def_sunrise}) the three internal masses are denoted by $m_1$, $m_2$ and $m_3$. 
The arbitrary scale $\mu$ is introduced to make the integral dimensionless.
$p^2$ denotes the momentum squared.
This variable plays an important role in our derivation and it is convenient
to introduce the notation
\bq
 t & = & p^2.
\eq
Where it is not essential we will suppress the dependence on the masses $m_i$ and the scale $\mu$ and simply write
$S( D, t)$ instead of $S( D, t, m_1^2, m_2^2, m_3^2, \mu^2)$.
In terms of Feynman parameters the two-loop integral is given by
\bq
\label{def_Feynman_integral}
 S\left( D, t\right)
 & = & 
 \Gamma\left(3-D\right)
 \left(\mu^2\right)^{3-D}
 \int\limits_{\sigma} \frac{{\cal U}^{3-\frac{3}{2}D}}{{\cal F}^{3-D}} \omega
\eq
with the two Feynman graph polynomials
\bq
 {\cal U} & = & x_1 x_2 + x_2 x_3 + x_3 x_1,
 \nonumber \\
 {\cal F} & = & - x_1 x_2 x_3 t
                + \left( x_1 m_1^2 + x_2 m_2^2 + x_3 m_3^2 \right) {\cal U}.
\eq
The differential two-form $\omega$ is given by
\bq
 \omega & = & x_1 dx_2 \wedge dx_3 - x_2 dx_1 \wedge dx_3 + x_3 dx_1 \wedge dx_2.
\eq
The integration is over
\bq
 \sigma & = & \left\{ \left[ x_1 : x_2 : x_3 \right] \in {\mathbb P}^2 | x_i \ge 0, i=1,2,3 \right\}.
\eq
In order to facilitate a comparison with results in the literature we remark that 
the definition of the sunrise integral in eq.~(\ref{def_sunrise}) is in Minkowski space. 
In a space with Euclidean signature one defines
the two-loop sunrise integral as
\bq
\lefteqn{
 S_{\mathrm{eucl}}\left( D, P^2, m_1^2, m_2^2, m_3^2, \mu^2 \right)
 = } & &
 \nonumber \\
 & &
 \left(\mu^2\right)^{3-D}
 \int \frac{d^DK_1}{\pi^{\frac{D}{2}}} \frac{d^DK_2}{\pi^{\frac{D}{2}}}
 \frac{1}{\left(K_1^2+m_1^2\right)\left(K_2^2+m_2^2\right)\left(\left(P-K_1-K_2\right)^2+m_3^2\right)}.
\eq
The momenta in Euclidean space are denoted by capital letters, while the ones in Minkowski space are denoted by lower case letters.
We have the relation
\bq
 S_{\mathrm{eucl}}\left( D, P^2, m_1^2, m_2^2, m_3^2, \mu^2 \right)
 & = &
 S\left( D, -p^2, m_1^2, m_2^2, m_3^2, \mu^2 \right).
\eq
Integration-by-parts identities can be used to derive relations between integrals with different powers of 
the propagators \cite{Tkachov:1981wb,Chetyrkin:1981qh}. 
Setting
\bq
 S_0 & = &  S\left( D, t, m_1^2, m_2^2, m_3^2, \mu^2, 1, 1, 1 \right) = S\left( D, t, m_1^2, m_2^2, m_3^2, \mu^2 \right),
 \nonumber \\
 S_1 & = &  S\left( D, t, m_1^2, m_2^2, m_3^2, \mu^2, 2, 1, 1 \right) = - \mu^2 \frac{\partial}{\partial m_1^2} S\left( D, t, m_1^2, m_2^2, m_3^2, \mu^2 \right),
 \nonumber \\
 S_2 & = &  S\left( D, t, m_1^2, m_2^2, m_3^2, \mu^2, 1, 2, 1 \right) = - \mu^2 \frac{\partial}{\partial m_2^2} S\left( D, t, m_1^2, m_2^2, m_3^2, \mu^2 \right),
 \nonumber \\
 S_3 & = &  S\left( D, t, m_1^2, m_2^2, m_3^2, \mu^2, 1, 1, 2 \right) = - \mu^2 \frac{\partial}{\partial m_3^2} S\left( D, t, m_1^2, m_2^2, m_3^2, \mu^2 \right),
\eq
where 
\bq
\lefteqn{
S\left( D, p^2, m_1^2, m_2^2, m_3^2, \mu^2, \nu_1, \nu_2, \nu_3 \right)
 = 
} & & 
 \\
 & & 
 \left(\mu^2\right)^{\nu_1+\nu_2+\nu_3-D}
 \int \frac{d^Dk_1}{i \pi^{\frac{D}{2}}} \frac{d^Dk_2}{i \pi^{\frac{D}{2}}}
 \frac{1}{\left(-k_1^2+m_1^2\right)^{\nu_1}\left(-k_2^2+m_2^2\right)^{\nu_2}\left(-\left(p-k_1-k_2\right)^2+m_3^2\right)^{\nu_3}}
 \nonumber
\eq
is the sunrise integral with arbitrary powers of the propagators,
one obtains in this way a set of four coupled first-order differential equations for the four quantities $S_0$, $S_1$, $S_2$ and $S_3$.
The set of first-order differential equations can be found in ref.~\cite{Caffo:1998du}.
Within the context of integration-by-parts identities the four quantities $S_0$, $S_1$, $S_2$ and $S_3$ are referred to as master integrals.
In this paper we will derive a single second-order differential equation for the quantity $S_0$.

It will be convenient to derive this equation in $D=2$ dimensions.
Dimensional recurrence relations \cite{Tarasov:1996br,Tarasov:1997kx,Baikov:1996iu,Lee:2009dh} can be used to obtain
the result in $D=4-2\eps$ dimensions.
Since the application of the dimensional recurrence relations involves a subtlety, we have addressed this issue in an appendix.
Working in $D=2$ dimensions has two advantages: First of all, the integral $S\left(2, t \right)$ is finite and does not require
regularisation.
The second advantage is given by the fact, that in $D=2$ dimensions eq.~(\ref{def_Feynman_integral}) reduces to
\bq
\label{def_Feynman_integral_dim_two}
 S\left( 2, t\right)
 & = & 
 \mu^2
 \int\limits_{\sigma} \frac{\omega}{{\cal F}}.
\eq
In $D=2$ dimensions the Feynman integral depends only on the second Symanzik polynomial ${\cal F}$, which occurs in the denominator, but not explicitly on the first
Symanzik polynomial ${\cal U}$.

Let us briefly discuss the simpler case of equal masses.
Setting $m_1=m_2=m_3=m$ it has been shown by Laporta and Remiddi \cite{Laporta:2004rb} that the system of four coupled first-order differential
equations reduces to a single second-order differential equation.
In $D=2$ dimensions this equation reads:
\bq
\label{diff_eq_equal_masses}
 \left[
  \frac{d^2}{dt^2}
 + \frac{\left( 3 t^2 - 20 t m^2 + 9 m^4 \right)}{t \left( t - m^2 \right) \left( t - 9 m^2 \right)} \frac{d}{dt}
 + \frac{t - 3 m^2}{t \left( t - m^2 \right) \left( t - 9 m^2 \right)}
 \right] S\left(2,t\right)
 & = & 
 \frac{-6 \mu^2}{t \left( t - m^2 \right) \left( t - 9 m^2 \right)}.
 \nonumber \\
\eq
In this paper we will show that this second-order differential equation is not an artifact of the special case $m_1=m_2=m_3=m$,
but has its origins in the interpretation of the zero set of ${\cal F}$ within algebraic geometry.

\section{Derivation of the second-order differential equation}
\label{sec:derivation}

In this section we show that the two-loop sunrise graph with arbitrary masses has a second order differential equation
similar to the one in eq.~(\ref{diff_eq_equal_masses}) in the case of equal masses. 
In this section we assume that all masses are positive and rational.

\subsection{Formalism}
\label{subsec:formalism}

Our starting point is eq.~(\ref{def_Feynman_integral_dim_two}):
\bq
 S\left( 2, t\right)
 & = & 
 \mu^2
 \int\limits_{\sigma} \frac{\omega}{{\cal F}}
 =
 \int\limits_{\sigma} \omega_t,
\eq
where we have set $\omega_t = \mu^2 \omega/{\cal F}$.
We interpret this Feynman integral as a period of a variation of a mixed Hodge structure (VMHS), varying with $t$. 
Although the fibres depend on the masses,
the rank of the VMHS is independent of them and so the differential equations should be of the same complexity in the unequal mass case compared to
the equal mass case.
We denote by ${\cal X}$ the set of points $\left(\left[ x_1 : x_2 : x_3 \right], t\right) \in {\mathbb P}^2 \times \Delta^\ast $, for which ${\cal F} = 0$.
Here, $\Delta^\ast$ is an open subset of ${\mathbb C}$.
We denote the fibre over $t$ by $X_t$.
The second Symanzik polynomial ${\cal F}$ is of degree three in the variables $x_1$, $x_2$ and $x_3$ and the zero set defines a family of elliptic curve
in ${\mathbb P}^2$ depending on $t$.
Then $\omega_t\in H^2(\mathbb P^2\text{\textbackslash} X_t)$ for all $t$, but $\sigma\notin H_2(\mathbb P^2\text{\textbackslash} X_t)$,
i.e. $\sigma$ is not a cycle, for two reasons.
Firstly, $\sigma$ intersects $X_t$ and secondly $\sigma$ has a boundary. The second fact being obvious, we now deal with the first.

\begin{Lemma}
\label{treshold_lemma}
Let $t_0:=(m_1+m_2+m_3)^2$ and $\C_{< t_0}$ be the complex numbers with the line $\{x\in \R\mid x\geq t_0\}$ removed. For any $t\in\C_{< t_0}$ the
chain of integration $\sigma$ intersects the graph hypersurface $X_t$ precisely in the three points $[1:0:0],\ [0:1:0]\ \text{and}\ [0:0:1]$.
\end{Lemma}
\noindent
{\bf Proof 1}\;
We have $\mathcal F=-t \ x_1 x_2 x_3+(m_1^2 x_1+m_2^2 x_2+m_3^2 x_3)(x_1 x_2+x_1 x_3+x_2 x_3)$. First observe that the boundary of $\sigma$ intersects $X_t$
precisely in the three points stated. We have to show that the intersection of $X_t$ with the inner points of $\sigma$ is the empty set for $t\in\C_{< t_0}$.
This is obvious for $t\in \C\text{\textbackslash}\ \R$. Now let $t=(m_1+m_2+m_3)^2-\delta$, with $\delta \in\R_{>0}$. We restrict to the
affine open $x_1=1$ and obtain the function
\[\mathcal F=-((m_1+m_2+m_3)^2-\delta ) x_2x_3+(m_1^2 +m_2^2 x_2+m_3^2 x_3)(x_2+x_3+x_2x_3).\]
We have to show that the equation
\[(m_1+m_2+m_3)^2-\delta =(m_1^2 +m_2^2 x_2+m_3^2 x_3)(\frac{1}{x_2}+\frac{1}{x_3}+1)\]
has no positive real solution. Now
$\varphi (x_2,x_3) :=(m_1^2 +m_2^2 x_2+m_3^2 x_3)(\frac{1}{x_2}+\frac{1}{x_3}+1)$ is a continuous function from
$U:=\R_{>0} \times\R_{>0}$ to $\R_{>0}$ which tends to infinity,
when $x_2$ or $x_3$ tend to zero or infinity. Hence the set $\{x\in U\mid \varphi(x)\leq C\} =: K\subset U$ is compact and $\varphi$
has it's global minimum on $K$.
We easily find the global minimum to be unique, namely the point $(x_2, x_3)=(\frac{m_1}{m_2},\frac{m_1}{m_3})$.
Now we have $\varphi (\frac{m_1}{m_2},\frac{m_1}{m_3})=(m_1+m_2+m_3)^2$ which proves the Lemma (the other two affine opens give the same solution). $\hfill\square$

In the following we will assume $t\in\C_{< t_0}$.
The differential equation which we derive will be valid in the region ${\mathbb C}_{< t_0}$.
Note that $p^2=t_0=(m_1+m_2+m_3)^2$ is the physical threshold. 
The two-loop sunrise integral for values of $p^2$ above the threshold can be obtained from the solution of the differential equation 
by analytic continuation with the help of Feynman's $i \eps$ prescription.

Now let $P\stackrel{\pi}{\longrightarrow}\mathbb P^2$ be the blow up of $\mathbb P^2$ in the three points of Lemma ~\ref{treshold_lemma}.
We denote the strict transform of $X_t$ by $Y_t$ and the strict transform of $\sigma$ again by $\sigma$.
In the particular example of the two-loop sunrise graph we are in the lucky situation that $X_t$ is isomorphic to $Y_t$ for generic $t$ -- both are
smooth elliptic curves. In $P$ we have $\sigma\cap Y_t=\emptyset$. Now let $B_0:=\{x_1 x_2 x_3=0\}\subset \mathbb P^2$ and $B$ it's total transform.
Clearly the boundary of $\sigma$ is contained in $B$. We now find
\[H_t:=H^2(P\ \text{\textbackslash}\ Y_t,\ B\ \text{\textbackslash}\ B\cap Y_t)\]
to be the right mixed Hodge structure, i.e. $\omega_t\in H^{2}(P\ \text{\textbackslash}\ Y_t, B\ \text{\textbackslash}\ B\cap Y_t)$ and
$\sigma\in H_2(P\ \text{\textbackslash}\ Y_t, B\ \text{\textbackslash}\ B\cap Y_t)$. Note that this is very similar to the work of Bloch, Esnault and
Kreimer \cite{Bloch:2006}. The convergent Feynman--Integral $S(2,t)$ is a period of $H^2(P\ \text{\textbackslash}\ Y_t,\ B\ \text{\textbackslash}\ B\cap Y_t)$.
We compute it's Picard--Fuchs equation.
\\

In the following we will denote a generic fibre by $X$, resp. $Y$, dropping the subscript $t$.
Since $Y$ is smooth we can apply the Gysin sequence which is a sequence of mixed Hodge structures and reads
\begin{gather*}
0\longrightarrow H^1(P\text{\textbackslash}\ Y)\longrightarrow H^0(Y) \longrightarrow H^2(P) \longrightarrow H^2(P\text{\textbackslash}\ Y)
\longrightarrow H^1(Y) \longrightarrow 0,
\end{gather*}
where we have used $H^1(P)=H^3(P)=0$. Indeed we have $H^k(P)=H^k(\mathbb P^2)$ for $k\neq 2$ and
$H^2(P)=\pi^{*}([w])\oplus \Z E_1\oplus \Z E_2\oplus \Z E_3$, where $w$ is a generator of $H^2(\mathbb P^2)$ and the $E_i$ correspond to the
exceptional divisors of the blowup (which are $\mathbb P^1s$). But $i_*$ maps $H^0(Y)$ isomorphically onto $\pi^{*}(w)$, such that we get
$H^1(P\text{\textbackslash}\ Y)=0$ and a short exact sequence
\begin{gather} \begin{CD}
0 @>>> \Z E_1\oplus \Z E_2\oplus \Z E_3 @>>> H^2(P\text{\textbackslash}\ Y) @>res>> H^1(Y) @>>> 0.
\label{gysin}
\end{CD}
\end{gather}

This sequence is split as a sequence of mixed Hodge structures via
\begin{gather}
\begin{CD}
H^2(P\text{\textbackslash} Y) @>{res}>> H^1(Y)\\
@A{\pi^{*}}AA		@VV{\cong}V \\
H^2(\mathbb P^2\text{\textbackslash} X) @>{res}>{\cong}> H^1(X).
\label{comm_diagram_homogeneous_part}
\end{CD}
\end{gather}

The elliptic curve $X$ has a unique holomorphic one--form (unique up to exact forms), which reads in Weierstrass normal coordinates
$\frac{dx}{y}$ if we restrict to the affine open $z=1$.
The Picard--Fuchs equation of $\frac{dx}{y}\in H^1(X)$ can easily be computed. We find the Picard--Fuchs operator
\bq
\label{pic_fuchs_op}
L^{(2)}=\frac{d^2}{d t^2} +a(t) \frac{d}{d t} +b(t),
\eq
with rational functions $a(t)$ and $b(t)$.\\
This is also the Picard--Fuchs operator of the Feynman form $\omega_t\in H^2(P\text{\textbackslash}\ Y)$ due to the splitting of sequence \eqref{gysin}
and the flatness of the system $\Z E_1\oplus \Z E_2\oplus \Z E_3$. So for any cycle $\xi$ in $H_2(P\text{\textbackslash}\ Y)$ we have
\[L^{(2)}(\int\limits_\xi \omega_t)=0.\]

Since the domain of integration $\sigma$ is not a cycle in $H_2\left(P\backslash Y\right)$, as explained above, we now pass to the relative setting.
There is the long exact sequence of relative cohomology
\begin{gather*}
0 \longrightarrow H^1(B\text{\textbackslash} B\cap Y) \longrightarrow H^2(P\text{\textbackslash} Y,\ B\text{\textbackslash} B\cap Y)
 \longrightarrow H^2(P\text{\textbackslash} Y) \longrightarrow H^2(B\text{\textbackslash} B\cap Y),
\end{gather*}
where we have used $H^1(P\text{\textbackslash} Y)=0$, as shown above. But now we find
\[B\cap Y=\{[0:-\frac{m_3}{m_2}:1],[-\frac{m_3}{m_1}:0:1],[1:-\frac{m_1}{m_2}:0]\}\cup \{p_1,p_2,p_3\},\]
where $p_i$ is a point on the exceptional divisor $E_i$.

\begin{figure}
\begin{center}
\begin{picture}(100,100)(0,0)
\Line(15,80)(85,80)
\GCirc(50,80){2}{1}
\Line(15,20)(85,20)
\GCirc(50,20){2}{1}
\Line(5,55)(25,15)
\GCirc(15,35){2}{1}
\Line(5,45)(25,85)
\GCirc(15,65){2}{1}
\Line(95,55)(75,15)
\GCirc(85,35){2}{1}
\Line(95,45)(75,85)
\GCirc(85,65){2}{1}
\end{picture}
\end{center}
\caption{
$B\text{\textbackslash}\ B\cap Y$.
The lines are ${\mathbb P}^1$s. The circles indicate, that one point is removed from each copy of ${\mathbb P}^1$.
}
\label{fig_B}
\end{figure}
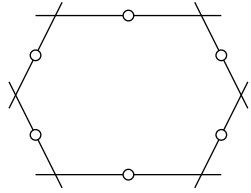

For $B\text{\textbackslash} B\cap Y$ we get the picture of figure~\ref{fig_B}.
Using the Mayer--Vietoris sequence we get
\begin{align*}
H^0(B\text{\textbackslash} B\cap Y)&=\Z,\\
H^1(B\text{\textbackslash} B\cap Y)&=\Z(-1) \text{,\ and} \\
H^k(B\text{\textbackslash} B\cap Y)&=0 \text{,\ for\ } k\neq 1,2.
\end{align*}
Summarising, we get the short exact sequence
\begin{gather}
\begin{CD}
0 @>>> \Z(-1) @>>> H^2(P\text{\textbackslash} Y,\ B\text{\textbackslash} B\cap Y) @>>> H^2(P\text{\textbackslash} Y) @>>> 0
\label{relative}
\end{CD}
\end{gather}
of mixed Hodge structures.

\begin{Prop}
The Picard--Fuchs operator of the Feynman integral $S(2,t)$ is\linebreak
$L:=\Bigl(\frac{d}{dt}-\frac{g'(t)}{g(t)}\Bigr) L^{(2)}$, where $g(t)$ is a rational function of $t$ with complex coefficients.
\end{Prop}
\noindent
{\bf Proof 2}\;
We have seen $L^{(2)}(\omega_t)=d\beta_t$, for some one--form $\beta_t$. This amounts to the identity
\[L^{(2)}(\int\limits_\sigma \omega_t)=\int\limits_\sigma d\beta_t=\int\limits_{\partial\sigma} \beta_t=:g(t).\]
From sequence $(\ref{relative})$ we know that the Picard Fuchs operator of $S(2,t)$ can have order three at most.
Obviously $L(S(2,t))=0$. 
It remains to show that $L^{(2)}(S(2,t))\neq 0$. 
Our explicit calculation shows that this is indeed the case. $\hfill\square$

From 
\bq
 L ( \int\limits_\sigma \omega_t ) & = & 0
\eq
we obtain immediately
\bq
 \left[ \frac{d^2}{d t^2} +a(t) \frac{d}{dt} +b(t) \right] S\left(2,t\right) & = & g(t).
\eq
This is the sought-after second-order differential equation. It remains to determine the coefficients $a(t)$ and $b(t)$, as well as the
inhomogeneous term $g(t)$.

\subsection{Calculation of the homogeneous part}
\label{subsec:homogeneous}

In this subsection we outline the calculation of the coefficients $a(t)$ and $b(t)$ in the homogeneous part of the differential equation.
From eq.~(\ref{comm_diagram_homogeneous_part}) it follows that it is sufficient to consider $H^1(X)$.
We recall that the variety $X$ is defined for fixed $t$ as the zero set in ${\mathbb P}^2$ of the second Symanzik polynomial ${\cal F}$:
\bq
 - x_1 x_2 x_3 t + \left( x_1 m_1^2 + x_2 m_2^2 + x_3 m_3^2 \right) \left( x_1 x_2 + x_2 x_3 + x_3 x_1 \right) & = & 0.
\eq
This polynomial is of degree $3$ in the variables $x_1$, $x_2$ and $x_3$ and defines an elliptic curve.
By an birational change of coordinates the defining equation can be brought into the Weierstrass normal form
\bq
 y^2 z - x^3 - a_2(t) x z^2 - a_3(t) z^3 & = & 0.
\eq
In the chart $z=1$ this reduces to
\bq
 y^2 - x^3 - a_2(t) x - a_3(t) & = & 0.
\eq
In this coordinates $H^1(X)$ is generated by
\bq
 \eta = \frac{dx}{y}
 & \mbox{and} &
 \dot{\eta} = \frac{d}{dt} \eta.
\eq
Since $H^1(X)$ is two-dimensional it follows that $\ddot{\eta}=\frac{d^2}{dt^2} \eta$ must be a linear combination of $\eta$ and $\dot{\eta}$.
In other words we must have a relation of the form
\bq
\label{eq_eta}
 \ddot{\eta} + a(t) \dot{\eta} + b(t) \eta & = & 0.
\eq
It is convenient to bring this equation onto a common denominator. Doing so and carrying out the derivatives with respect to $t$ we have
\bq
 \eta & = & \left( x^3 + a_2 x + a_3 \right)^2 \frac{dx}{y^5},
 \nonumber \\
 \dot{\eta} & = & - \frac{1}{2} \left( \dot{a}_2 x + \dot{a}_3 \right) \left( x^3 + a_2 x + a_3 \right) \frac{dx}{y^5},
 \nonumber \\
 \ddot{\eta} & = & \left[ - \frac{1}{2} \left( \ddot{a}_2 x + \ddot{a}_3 \right) \left( x^3 + a_2 x + a_3 \right)
                          + \frac{3}{4} \left( \dot{a}_2 x + \dot{a}_3 \right)^2 \right] \frac{dx}{y^5}.
\eq
The numerator of eq.~(\ref{eq_eta}) is then a polynomial of degree $6$ in the single variable $x$.
Since we work in $H^1(X)$, we can simplify the expression by adding an exact form
\bq
 d \left( \frac{x^n}{y^3} \right)
 & = & x^{n-1} \left[ \left(n-\frac{9}{2}\right)x^3 + \left( n-\frac{3}{2} \right) a_2 x + n a_3 \right] \frac{dx}{y^5}.
\eq
This allows us to reduce the numerator polynomial from degree six to a linear polynomial.
The two coefficients of this linear polynomial have to vanish, on account of eq.~(\ref{eq_eta}).
We obtain therefore two equations for the two unknown parameters $a(t)$ and $b(t)$. 
Solving for $a(t)$ and $b(t)$ we find
\bq
 a(t) = \frac{p_1(t)}{p_0(t)},
 & &
 b(t) = \frac{p_2(t)}{p_0(t)},
\eq
with
\bq
\label{res_p012}
 p_1(t) & = & 
  9 t^6
  - 32 M_{100} t^5
  + \left( 37 M_{200} + 70 M_{110} \right) t^4
  - \left( 8 M_{300} + 56 M_{210} + 144 M_{111} \right) t^3
 \nonumber \\
 & &
  - \left( 13 M_{400} - 36 M_{310} + 46 M_{220} - 124 M_{211} \right) t^2
 \nonumber \\
 & &
  - \left( -8 M_{500} + 24 M_{410} - 16 M_{320} - 96 M_{311} + 144 M_{221} \right) t
 \nonumber \\
 & &
  - \left( M_{600} - 6 M_{510} + 15 M_{420} - 20 M_{330} + 18 M_{411} - 12 M_{321} - 6 M_{222} \right),
 \nonumber \\
 p_2(t) & = &
  3 t^5
  - 7 M_{100} t^4
  + \left( 2 M_{200} + 16 M_{110} \right) t^3
  + \left( 6 M_{300} - 14 M_{210} \right) t^2
 \nonumber \\
 & &
  - \left( 5 M_{400} - 8 M_{310} + 6 M_{220} - 8 M_{211} \right) t
  + \left( M_{500} - 3 M_{410} + 2 M_{320} + 8 M_{311} - 10 M_{221} \right),
 \nonumber \\
 p_0(t) & = &
  t \left[ t - \left( m_1+m_2+m_3\right)^2 \right]
    \left[ t - \left( -m_1+m_2+m_3\right)^2 \right]
    \left[ t - \left( m_1-m_2+m_3\right)^2 \right]
 \nonumber \\
 & &
    \left[ t - \left( m_1+m_2-m_3\right)^2 \right]
    \left[ 3 t^2 - 2 M_{100} t - M_{200} + 2 M_{110} \right].
\eq
In order to present the result in a compact form we have introduced 
the monomial symmetric polynomials $M_{\lambda_1 \lambda_2 \lambda_3}$ in the variables $m_1^2$, $m_2^2$ and $m_3^2$.
These are defined by
\bq
 M_{\lambda_1 \lambda_2 \lambda_3} & = &
 \sum\limits_{\sigma} \left( m_1^2 \right)^{\sigma\left(\lambda_1\right)} \left( m_2^2 \right)^{\sigma\left(\lambda_2\right)} \left( m_3^2 \right)^{\sigma\left(\lambda_3\right)},
\eq
where the sum is over all distinct permutations of $\left(\lambda_1,\lambda_2,\lambda_3\right)$.
A few examples are
\bq
 M_{100} & = & m_1^2 + m_2^2 + m_3^2,
 \nonumber \\
 M_{111} & = & m_1^2 m_2^2 m_3^2,
 \nonumber \\
 M_{210} & = & m_1^4 m_2^2 + m_2^4 m_3^2 + m_3^4 m_1^2 + m_2^4 m_1^2 + m_3^4 m_2^2 + m_1^4 m_3^2.
\eq

\subsection{Calculation of the inhomogeneous part}
\label{subsec:inhomogeneous}

In this subsection we compute the inhomogeneous part $g(t)$.
From proposition 2
we first seek a one-form $\beta_t$, such that
\bq
\label{eq_beta}
 d \beta_t & = & L^{(2)} \left( \omega_t \right)
\eq
We make the ansatz \cite{Griffiths:1969}
\bq
 \beta_t & = & 
 \frac{1}{p_0(t) {\cal F}^2} 
 \left[ \left( x_2 q_3 - x_3 q_2 \right) dx_1 + \left( x_3 q_1 - x_1 q_3 \right) dx_2 + \left( x_1 q_2 - x_2 q_1 \right) dx_3 \right],
\eq
where $q_1$, $q_2$ and $q_3$ are polynomials of degree $4$ in the variables $x_1$, $x_2$ and $x_3$.
The most general form is
\bq
\lefteqn{
 q_i = } & & \nonumber \\
 & &
 c^{(i)}_{400} x_1^4 + c^{(i)}_{040} x_2^4 + c^{(i)}_{004} x_3^4
 + 
 c^{(i)}_{310} x_1^3 x_2 + c^{(i)}_{301} x_1^3 x_3 + c^{(i)}_{130} x_1 x_2^3 + c^{(i)}_{103} x_1 x_3^3 + c^{(i)}_{031} x_2^3 x_3 + c^{(i)}_{013} x_2 x_3^3
 \nonumber \\
 & &
 +
 c^{(i)}_{211} x_1^2 x_2 x_3 + c^{(i)}_{121} x_1 x_2^2 x_3 + c^{(i)}_{112} x_1 x_2 x_3^2
 +
 c^{(i)}_{220} x_1^2 x_2^2 + c^{(i)}_{202} x_1^2 x_3^2 + c^{(i)}_{022} x_2^2 x_3^2.
\eq   
We would like $\beta_t$ to be finite on the boundary $\partial \sigma$. This implies
\bq
 c^{(1)}_{040} = c^{(1)}_{004} = c^{(2)}_{400} = c^{(2)}_{004} = c^{(3)}_{400} = c^{(3)}_{040} = 0.
\eq
The remaining 39 coefficients $c^{(i)}_{jkl}$ are found by solving the linear system of equations obtained from inserting the
ansatz into eq.~(\ref{eq_beta}).
The solution of this linear system is not unique, corresponding to the fact that $\beta_t$ can be changed by a closed one-form.
The solutions for the coefficients $c^{(i)}_{jkl}$ are rather lengthy and not listed here.
In the next step we integrate $\beta_t$ along the boundary $\partial \sigma$ to get $g(t)$:
\bq
 g(t) & = & \int\limits_{\partial \sigma} \beta_t.
\eq
Note that the integration is in the blow-up $P$ of ${\mathbb P}^2$. 
We obtain
\bq
 g(t) & = & \frac{p_3(t)}{p_0(t)},
\eq
with
\bq
\label{res_p3}
 p_3(t) & = & 
 -18 t^4
 + 24 M_{100} t^3
 + \left( 4 M_{200} - 40 M_{110} \right) t^2
 + \left( - 8 M_{300} + 8 M_{210} + 48 M_{111} \right) t
 \nonumber \\
 & & 
 + \left( - 2 M_{400} + 8 M_{310} - 12 M_{220} - 8 M_{211} \right)
 + 2 c\left(t,m_1,m_2,m_3\right)  \ln \frac{m_1^2}{\mu^2}
 \nonumber \\
 & &
 + 2 c\left(t,m_2,m_3,m_1\right)  \ln \frac{m_2^2}{\mu^2}
 + 2 c\left(t,m_3,m_1,m_2\right)  \ln \frac{m_3^2}{\mu^2}
\eq
and
\bq
\lefteqn{
c\left(t,m_1,m_2,m_3\right) = } & &
 \nonumber \\
 & &
 \left( -2 m_1^2 + m_2^2 + m_3^2 \right) t^3
 + \left( 6 m_1^4 - 3 m_2^4 - 3 m_3^4 - 7 m_1^2 m_2^2 - 7 m_1^2 m_3^2 + 14 m_2^2 m_3^2 \right) t^2
 \nonumber \\
 & &
 + \left( -6 m_1^6 + 3 m_2^6 + 3 m_3^6 + 11 m_1^4 m_2^2 + 11 m_1^4 m_3^2 - 8 m_1^2 m_2^4 - 8 m_1^2 m_3^4 - 3 m_2^4 m_3^2 - 3 m_2^2 m_3^4 \right) t
 \nonumber \\
 & & 
 + \left( 2 m_1^8 - m_2^8 - m_3^8 - 5 m_1^6 m_2^2 - 5 m_1^6 m_3^2 + m_1^2 m_2^6 + m_1^2 m_3^6 + 4 m_2^6 m_3^2 + 4 m_2^2 m_3^6 
 \right. \nonumber \\
 & & \left.
        + 3 m_1^4 m_2^4 + 3 m_1^4 m_3^4 - 6 m_2^4 m_3^4 
        + 2 m_1^4 m_2^2 m_3^2 - m_1^2 m_2^4 m_3^2 - m_1^2 m_2^2 m_3^4 \right).
\eq
The coefficients $c(t,m_i,m_j,m_k)$ of the logarithms of the masses vanish for equal masses.

\subsection{Final result}
\label{subsec:result}

In this subsection we summarise our results.
The two-loop sunrise integral $S(2,t)$ with unequal masses satisfies a second-order differential equation.
This second order differential equation is given by
\bq
\label{res_final}
 \left[ \frac{d^2}{d t^2} + \frac{p_1(t)}{p_0(t)} \frac{d}{dt} + \frac{p_2(t)}{p_0(t)}  \right] S\left(2,t\right) & = & \mu^2 \frac{p_3(t)}{p_0(t)},
\eq
where $p_0(t)$, $p_1(t)$, $p_2(t)$ and $p_3(t)$ are polynomials in $t$.
The polynomials $p_0(t)$, $p_1(t)$ and $p_2(t)$ are defined in eq.~(\ref{res_p012}),
while $p_3(t)$ is given in eq.~(\ref{res_p3}).
In the special case of equal masses $m_1=m_2=m_3=m$, eq.~(\ref{res_final}) reduces to eq.~(\ref{diff_eq_equal_masses}).

\section{Conclusions}
\label{sec:conclusions}

In this paper we have shown that the two-loop sunrise integral in two dimensions with three arbitrary masses
has a second-order differential equation.
This differential equation is derived in a new way:
We view the Feynman integral as a period of a variation
of a mixed Hodge structure, where the variation is with respect to the external momentum squared.
We expect our technique to be applicable to other Feynman integrals as well, in the following sense:
In seeking a differential equation with respect to an external invariant for an unknown Feynman integral
we view the Feynman integral as a period of a variation of a mixed Hodge structure.
The fibre will be the complement of an algebraic variety.
The algebraic variety is given by the zero set of the Feynman polynomial ${\cal F}$
and depends therefore on the specific Feynman integral under consideration.
In the specific case of the two-loop sunrise integral the algebraic variety is an elliptic curve,
while for other integrals of interest we expect to encounter different algebraic varieties.

\subsection*{Acknowledgements}

We would like to thank the SFB/TR 45 ``Periods, moduli spaces and arithmetic of algebraic varieties'' for its support.
R.Z. acknowledges in addition the support of the research centre ``Elementary Forces and Mathematical Foundations''.
We are grateful to S. Groote and J. K\"orner for checking numerically with the methods of ref.~\cite{Groote:2005ay} 
the correctness of our results.


\begin{appendix}

\section{Dimensional recurrence relations}

Although the focus of this article lies on the two-loop sunrise integral in two dimensions, we would
like to give in this appendix some indications how the integral in two dimensions can be related to the
one in $D=4-2\eps$ dimensions.
In the unequal mass case there are some subtleties, which are discussed in this appendix.

Dimensional recurrence relations can be used to relate the two-loop sunrise graph in $D=4-2\eps$ dimensions
to the result in $D=2-2\eps$ dimensions.
Let us introduce an operator ${\bf i}^+$, which raises the power of the propagator $i$ by one, e.g.
\bq
 {\bf 1}^+ S(D,t,m_1^2,m_2^2,m_3^2,\nu_1,\nu_2,\nu_3,\mu^2) 
 & = & 
 S(D,t,m_1^2,m_2^2,m_3^2,\nu_1+1,\nu_2,\nu_3,\mu^2).
\eq
The integral $S(D,t,m_1^2,m_2^2,m_3^2,\nu_1,\nu_2,\nu_3,\mu^2)$ is defined by
\bq
\label{sunrise_arbitrary_props}
\lefteqn{
 S\left( D, p^2, m_1^2, m_2^2, m_3^2, \nu_1,\nu_2,\nu_3,\mu^2 \right)
 = } & &
 \\
 & = &
 \left(\mu^2\right)^{\nu-D}
 \int \frac{d^Dk_1}{i \pi^{\frac{D}{2}}} \frac{d^Dk_2}{i \pi^{\frac{D}{2}}}
 \frac{1}{\left(-k_1^2+m_1^2\right)^{\nu_1}\left(-k_2^2+m_2^2\right)^{\nu_2}\left(-\left(p-k_1-k_2\right)^2+m_3^2\right)^{\nu_3}}
 \nonumber \\
 & = &
 \frac{\Gamma\left(\nu-D\right)}{\Gamma\left(\nu_1\right)\Gamma\left(\nu_2\right)\Gamma\left(\nu_3\right)}
 \left(\mu^2\right)^{\nu-D}
 \int\limits_{\sigma} x_1^{\nu_1-1}x_2^{\nu_2-1}x_3^{\nu_3-1} \frac{{\cal U}^{\nu-\frac{3}{2}D}}{{\cal F}^{\nu-D}} \omega,
 \nonumber
\eq
where we have set $\nu=\nu_1+\nu_2+\nu_3$.

The starting point for the dimensional recurrence relation is given by \cite{Tarasov:1996br,Tarasov:1997kx}
\bq
\label{dim_shift_eq}
 S(D-2,t,m_1^2,m_2^2,m_3^2,\nu_1,\nu_2,\nu_3,\mu^2)
 = 
 {\cal U}\left(\nu_1 {\bf 1}^+, \nu_2 {\bf 2}^+, \nu_3 {\bf 3}^+\right)
 S(D,t,m_1^2,m_2^2,m_3^2,\nu_1,\nu_2,\nu_3,\mu^2).
 \nonumber \\
\eq
Eq.~(\ref{dim_shift_eq}) is easily verified by writing out the right-hand side:
\bq
\label{helper_1}
\lefteqn{
 {\cal U}\left(\nu_1 {\bf 1}^+, \nu_2 {\bf 2}^+, \nu_3 {\bf 3}^+\right)
 S(D,t,m_1^2,m_2^2,m_3^2,\nu_1,\nu_2,\nu_3,\mu^2)
= } & & 
 \nonumber \\
 & = & 
 \nu_1 \nu_2 S(D,t,m_1^2,m_2^2,m_3^2,\nu_1+1,\nu_2+1,\nu_3,\mu^2)
 +
 \nu_2 \nu_3 S(D,t,m_1^2,m_2^2,m_3^2,\nu_1,\nu_2+1,\nu_3+1,\mu^2)
 \nonumber \\
 & &
 +
 \nu_3 \nu_1 S(D,t,m_1^2,m_2^2,m_3^2,\nu_1+1,\nu_2,\nu_3+1,\mu^2).
\eq
Using eq.~(\ref{sunrise_arbitrary_props}) we can express
\bq
\lefteqn{
 \nu_1 \nu_2 S(D,t,m_1^2,m_2^2,m_3^2,\nu_1+1,\nu_2+1,\nu_3,\mu^2)
 = } & &
 \nonumber \\
 & & 
 \frac{\Gamma\left(\nu+2-D\right)}{\Gamma\left(\nu_1\right)\Gamma\left(\nu_2\right)\Gamma\left(\nu_3\right)}
 \left(\mu^2\right)^{\nu+2-D}
 \int\limits_{\sigma} x_1^{\nu_1}x_2^{\nu_2}x_3^{\nu_3-1}
 \frac{{\cal U}^{\nu+2-\frac{3}{2}D}}{{\cal F}^{\nu+2-D}} \omega,
\eq
and similar for the two other integrals on the right-hand side of eq.~(\ref{helper_1}).
Adding up the three integrals we find
\bq
\lefteqn{
 {\cal U}\left(\nu_1 {\bf 1}^+, \nu_2 {\bf 2}^+, \nu_3 {\bf 3}^+\right)
 S(D,t,m_1^2,m_2^2,m_3^2,\nu_1,\nu_2,\nu_3,\mu^2)
= } & & 
 \nonumber \\
 & = & 
 \frac{\Gamma\left(\nu+2-D\right)}{\Gamma\left(\nu_1\right)\Gamma\left(\nu_2\right)\Gamma\left(\nu_3\right)}
 \left(\mu^2\right)^{\nu+2-D}
 \int\limits_{\sigma} x_1^{\nu_1-1}x_2^{\nu_2-1}x_3^{\nu_3-1}
 \frac{{\cal U}^{\nu+3-\frac{3}{2}D}}{{\cal F}^{\nu+2-D}} \omega,
\eq
which is just $S(D-2,t,m_1^2,m_2^2,m_3^2,\nu_1,\nu_2,\nu_3,\mu^2)$.

Integration-by-parts identities \cite{Tkachov:1981wb,Chetyrkin:1981qh,Studerus:2009ye} are of the form
\bq
 \left(\mu^2\right)^{\nu-D}
 \int \frac{d^Dk_1}{i \pi^{\frac{D}{2}}} \frac{d^Dk_2}{i \pi^{\frac{D}{2}}}
 \frac{\partial}{\partial k_i^\mu}
 \frac{q^\mu}{\left(-k_1^2+m_1^2\right)^{\nu_1}\left(-k_2^2+m_2^2\right)^{\nu_2}\left(-k_3^2+m_3^2\right)^{\nu_3}}
 & = & 0,
\eq
with $k_i \in \{ k_1, k_2 \}$, $q \in \{ p, k_1, k_2 \}$ and $k_3=p-k_1-k_2$.
Integration-by-parts identities can be used to derive relations between integrals with the same topology but
different powers of the propagators and integrals with simpler topologies.
Using integration-by-parts identities 
we express the right hand side of eq.~(\ref{dim_shift_eq})
in terms of the master integrals $S_0$, $S_1$, $S_2$, $S_3$ and simpler integrals,
which are products of one-loop tadpole integrals.
The one-loop tapole integral is given by
\bq
 T\left( D, m^2, \mu^2 \right) 
 & = &
 \left(\mu^2\right)^{1-\frac{D}{2}}
 \int \frac{d^Dk}{i \pi^{\frac{D}{2}}} 
 \frac{1}{\left(-k^2+m^2\right)}
 =
 \Gamma\left(1-\frac{D}{2}\right)
 \left( \frac{m^2}{\mu^2} \right)^{\frac{D}{2}-1}.
\eq
Inverting a linear system of equations we can express the $D$ dimensional integrals
in terms of the $(D-2)$ dimensional integrals.
Since the involved expressions are rather lengthy, this is usually done by computer algebra.
Specialising to $D=4-2\eps$ we obtain
the relation
\bq
\label{eq_dim_recurrence_I}
 S\left(4-2\eps, t \right) & = &
 c_0\left(\eps,t\right) S\left(2-2\eps, t \right)
 +
 c_1\left(\eps,t\right) S_1\left(2-2\eps, t \right)
 +
 c_2\left(\eps,t\right) S_2\left(2-2\eps, t \right)
 \nonumber \\
 & &
 +
 c_3\left(\eps,t\right) S_3\left(2-2\eps, t \right)
 +
 r\left(\eps,t\right).
\eq
The dependence of the coefficients $c_0$, $c_1$, $c_2$, $c_3$ and $r$ on the masses $m_1$, $m_2$ and $m_3$ is not shown
explicitly.
All quantities in eq.~(\ref{eq_dim_recurrence_I}) can be viewed as a Laurent series in $\eps$.
The Laurent series of $S(4-2\eps,t)$ starts at $1/\eps^2$ and one is usually interested in the pole terms and in the
$\eps^0$-term.
The quantities $S(2-2\eps, t)$, $S_1(2-2\eps, t)$, $S_2(2-2\eps, t)$ and $S_3(2-2\eps, t)$ are finite, and their
Laurent series start at $\eps^0$.
If it is the case that also the Laurent series of the coefficients $c_0$, $c_1$, $c_2$ and $c_3$ start at $\eps^0$,
eq.~(\ref{eq_dim_recurrence_I}) would reduce to
\bq
\label{eq_dim_recurrence_II}
 S\left(4-2\eps, t \right) & = &
 c_0(0,t) S\left(2, t \right)
 +
 c_1(0,t) S_1\left(2, t \right)
 +
 c_2(0,t) S_2\left(2, t \right)
 +
 c_3(0,t) S_3\left(2, t \right)
 \nonumber \\
 & &
 +
 r\left(\eps,t\right)
 + 
 {\cal O}\left(\eps\right),
\eq
and it would be sufficient to know $S(2, t)$, $S_1(2, t)$, $S_2(2, t)$ and $S_3(2, t)$ in order to determine the $\eps^0$-term
of $S(4-2\eps, t)$.
The ultraviolet poles of $S(4-2\eps, t)$ would then be entirely given by the tadpole contributions contained in the coefficient $r$.
Unfortunately it turns out that for this choice of master integrals the coefficients $c_0$, $c_1$, $c_2$ and $c_3$
contain spurious poles in $\eps$.
In order to avoid the computation of the $\eps^1$-terms of the master integrals it is advantageous to choose a different basis
for the master integrals.
A basis which does not lead to spurious poles is given by
\bq
 I_1\left(D,t\right) & = & S\left(D,t,m_1^2,m_2^2,m_3^2,\mu^2\right),
 \\
 I_2\left(D,t\right) & = & \mu^2 \frac{d}{dt} I_1\left(D,t\right),
 \nonumber \\
 I_3\left(D,t\right) & = & 
  \left(\mu^2\right)^{2-D}
 \int \frac{d^Dk_1}{i \pi^{\frac{D}{2}}} \frac{d^Dk_2}{i \pi^{\frac{D}{2}}}
 \frac{2p \cdot k_1}{\left(-k_1^2+m_1^2\right) \left(-k_2^2+m_2^2\right) \left(-\left(p-k_1-k_2\right)^2+m_3^2\right)},
 \nonumber \\
 I_4\left(D,t\right) & = & 
  \left(\mu^2\right)^{2-D}
 \int \frac{d^Dk_1}{i \pi^{\frac{D}{2}}} \frac{d^Dk_2}{i \pi^{\frac{D}{2}}}
 \frac{2p \cdot k_2}{\left(-k_1^2+m_1^2\right) \left(-k_2^2+m_2^2\right) \left(-\left(p-k_1-k_2\right)^2+m_3^2\right)}.
 \nonumber
\eq
The second-order differential equation can be used to determine $I_1(2,t)$ and $I_2(2,t)$.
In addition, the integrals $I_3(2,t)$ and $I_4(2,t)$ are required.
These are given by
\bq
 I_3\left(2,t\right)
 & = & 2 p^2 \int\limits_{\sigma} \frac{x_2 x_3 \omega}{{\cal U} {\cal F}},
 \nonumber \\
 I_4\left(2,t\right)
 & = & 2 p^2 \int\limits_{\sigma} \frac{x_1 x_3 \omega}{{\cal U} {\cal F}}.
\eq
Only one integral of these two needs to be known, the other follows then from the substitution 
$m_1 \leftrightarrow m_2$.
We can also derive a differential equation for $I_3$ (or $I_4$) with the methods of this article.
The actual calculations will be slightly more complicated due to the extra factor $x_2 x_3 /{\cal U}$
(or $x_1 x_3 /{\cal U}$).
The integrals $I_3$ and $I_4$ will be treated in detail in a future publication.

\end{appendix}

\bibliography{/home/stefanw/notes/biblio}
\bibliographystyle{/home/stefanw/latex-style/h-physrev5}

\end{document}